\documentclass[manuscript]{acmart}

\AtBeginDocument{%
  \providecommand\BibTeX{{%
    \normalfont B\kern-0.5em{\scshape i\kern-0.25em b}\kern-0.8em\TeX}}}

\acmYear{2024}\copyrightyear{2024}
\setcopyright{rightsretained}
\acmConference[ACM FAccT '24]{ACM Conference on Fairness, Accountability, and Transparency}{June 3--6, 2024}{Rio de Janeiro, Brazil}
\acmBooktitle{ACM Conference on Fairness, Accountability, and Transparency (ACM FAccT '24), June 3--6, 2024, Rio de Janeiro, Brazil}
\acmDOI{10.1145/3630106.3659002}
\acmISBN{979-8-4007-0450-5/24/06}

\begin{document}

\title[Real Risks of Fake Data]{Real Risks of Fake Data: Synthetic Data, Diversity-Washing and Consent Circumvention}

\author{Cedric Deslandes Whitney}
\email{cedricwhitney@berkeley.edu}
\affiliation{%
  \institution{University of California, Berkeley}
  \city{Berkeley}
  \state{California}
  \country{USA}
  \postcode{94609}
}

\author{Justin Norman}
\email{justin.norman@berkeley.edu}

\affiliation{%
  \institution{University of California, Berkeley}
  \city{Berkeley}
  \state{California}
  \country{USA}
  \postcode{94609}
}

\renewcommand{\shortauthors}{Whitney and Norman}



\begin{abstract}
Machine learning systems require representations of the real world for training and testing - they require data, and lots of it. Collecting data at scale has logistical and ethical challenges, and synthetic data promises a solution to these challenges. Instead of needing to collect photos of real people’s faces to train a facial recognition system, a model creator could create and use photo-realistic, synthetic faces. The comparative ease of generating this synthetic data rather than relying on collecting data has made it a common practice. We present two key risks of using synthetic data in model development. First, we detail the high risk of false confidence when using synthetic data to increase dataset diversity and representation. We base this in the examination of a real world use-case of synthetic data, where synthetic datasets were generated for an evaluation of facial recognition technology. Second, we examine how using synthetic data risks circumventing consent for data usage. We illustrate this by considering the importance of consent to the U.S. Federal Trade Commission’s regulation of data collection and affected models. Finally, we discuss how these two risks exemplify how synthetic data complicates existing governance and ethical practice; by decoupling data from those it impacts, synthetic data is prone to consolidating power away those most impacted by algorithmically-mediated harm.
\end{abstract}

\maketitle

\section{Introduction}~\label{Intro}
Facial recognition technology (FRT) has become commonplace, used from flight check-in at airports to police crowd-monitoring. Bias in FRT models has resulted in mis-identification and expanded surveillance, causing unjust incarceration and other discriminatory outcomes. Attempts to solve these issues by increasing the accuracy of FRT run headfirst into problems; for a machine learning-based computer vision system to be considered robust enough for a given real-world task, it must “generalize” to images that vary widely in quality and domain (image granularity, race, age, gender, background, head pose, hats, glasses, etc.). Datasets with this level of granular design and annotation, that are also large enough for use in deep learning, are nearly impossible to find due to logistical and ethical concerns. As a result, researchers have turned to synthetic data generation, where data is generated to resemble something without being a representation of an instance of it --- a drawing of a generic face as compared to a photograph of a real person. Synthetic data has been used to augment existing datasets and create new datasets for better training and evaluation of FRT models. Logistical and ethical challenges to data collection exist outside of FRT, and synthetic data usage has become commonplace across machine learning, from computer vision to large language models. This paper examines two key risks of using synthetic data.

Synthetic data is fundamentally useful where real data is not fit to task, necessitating that synthetic data must be both similar enough to be meaningful, but different enough to mitigate the reason the real data is not usable~\cite{jordonSyntheticDataWhat2022}. Jordan et al. propose three attributes of synthetic data that must be met for it to function in lieu of real data: utility, fidelity, and privacy. This paper focuses on facial recognition because it clearly articulates the risks of synthetic data, inherently forcing trade-offs between these attributes. There is high difficulty in making a picture of a face private but still usable as training data (privacy vs. utility) --- a face which has been obscured to the point where an identity could not be gleaned is less useful~\cite{chamikaraPrivacyPreservingFace2020}. Achieving fidelity in facial datasets, a measure of how well synthetic data matches the real world, is also saliently difficult in facial recognition use cases, as we examine below.

The first risk we focus on is the \textit{high risk of false confidence} in the ability of synthetic datasets to mitigate bias in data distribution and representation. We demonstrate this through the real-world example of using synthetic data for a facial recognition model evaluation. This paper was motivated by the realization of the under-explored risks of synthetic data while conducting the evaluation, and we present it both to provide an example of how synthetic data is used and to detail the concerns that conducting it made apparent. In brief, synthetic data offers a way of diversifying datasets, but diversity in real-world faces often follows from cultural practices that are qualitative and meaning-laden rather than quantitative. Creating a synthetic dataset or adding synthetic data to existing datasets in an attempt to diversify that dataset runs the risk of \textit{diversity-washing} --- appearing to resolve valid criticism regarding a dataset’s distribution and representation but in a way that is superficial. As a result, using synthetic data risks legitimizing technologies such as FRT despite potentially continuing to propagate bias by achieving false fidelity.

The second risk we examine is how using synthetic data risks \textit{circumventing consent for data usage}, illustrating the impacts by considering the importance of consent to the U.S. Federal Trade Commission’s regulation of data collection and affected models. Synthetic data provides an avenue for model developers to side-step thorny issues around collecting large-scale representative facial datasets. Proper consent to data usage is foundational to the privacy enforcement tools that the FTC has used to require companies delete ML models trained on improperly collected data, a key regulatory hurdle to improper data collection and resulting harmful model deployment. Using synthetic data risks circumventing and obfuscating consent, thus complicating deterrence and enforcement.

This paper proceeds as follows: We begin in Section~\ref{RW} by summarizing related prior work. We first examine work on datasets and representation, before discussing participation and consent and power over data and models. Finally, we discuss synthetic data and its use. We then proceed to the two titular risks of synthetic data that this paper focuses on --- diversity-washing (Section~\ref{R1}) and the circumvention of consent (Section~\ref{R2}). We draw upon two real-world examples: a facial recognition evaluation task conducted using synthetic data, and the FTC’s enforcement actions against models trained on deceptively collected data to illustrate these risks. Finally, in Section~\ref{Dis} we expand upon how these two risks are examples of irresponsible use of synthetic data: consolidating power in the hands of model creators, and decoupling data from those it represents and those who are harmed by its improper use. It is our intention for this research to contribute to the field by presenting tangible examples and background for the challenges inherent in responsible use of synthetic data, thus laying foundations for further work and debate. We call for future work to examine the breadth and usage of synthetic data and to work towards both mitigating synthetic data’s risks and enabling its potential for participatory empowerment.

\section{Related Work}~\label{RW}
In this section, first, we focus on the datasets that underpin machine learning systems, and detail how that work treats the specific issue of data distribution and representation in those datasets (Section~\ref{RW:DDR}). Next, we discuss prior work on participatory governance, consent and data privacy, and attempts to capture some power over dataset creation and usage by those most affected (Section~\ref{RW:PCR}). Finally, we provide a summary of work detailing what synthetic data is and how it is used (Section~\ref{RW:SD}). 

\subsection{Datasets, Diversity and Representation}~\label{RW:DDR}
Datasets are the hidden infrastructure behind machine learning, most visible when the systems built on them break~\cite{jacksonRethinkingRepair2014}. Models developed and dependent on these large datasets can lead to biased and harmful effects, with models used for bureaucratic categorization in particular having a long history of harm~\cite{bullockArtificialIntelligenceDiscretion2019,alkhatibStreetLevelAlgorithmsTheory2019,spadeNormalLifeAdministrative2015a}. The collection of data is then frequently the starting point for ML-disseminated discrimination and bias in domains such as hiring~\cite{raghavanMitigatingBiasAlgorithmic2020}, advertising~\cite{leeAlgorithmicBiasDetection2019}, pricing~\cite{wuImpactAlgorithmicPrice2022}, the application of law, and government allocation of resources~\cite{abebeMechanismDesignSocial2018}; as well as being vital for identifying and enforcing against discrimination~\cite{andrusDemographicReliantAlgorithmicFairness2022}. The stakes of responsible dataset development, then, are high, and we build on critical previous work~\cite{paulladaDataItsDis2021,pengMitigatingDatasetHarms2021,hutchinsonAccountabilityMachineLearning2021} in focusing on the ways that dataset creators have significant impact on the harms that occur downstream via their development, usage and deployment~\cite{khanSubjectsStagesAI2022a}. More narrowly, we hope to bring focus to important risks present when synthetic data is used in the process of creating and using datasets in machine learning development.

Datasets used for facial recognition models, where the goal of the model is matching an image or video of a face to an identity, have received much scrutiny --- specifically for violating privacy~\cite{harveyResearchersGoneWild2021,birhaneLargeImageDatasets2021}. An analysis by Crawford \& Paglen~\cite{crawfordExcavatingAIPolitics2021} of ImageNet, a frequently-used large dataset, demonstrated active labeling of faces with offensive and derogatory classifications, and Prabhu \& Birhane~\cite{birhaneLargeImageDatasets2021} make the point that beyond obvious privacy harms such as blackmail, the creation of one of these datasets causes similar datasets to propagate. The recent identification of CSAM material in the popular LAION dataset is an example of this~\cite{thielIdentifyingEliminatingCSAM2023}. In the case of synthetic data, where the data is frequently derivative of previously collected data (as expanded upon below in Section~\ref{RW:SD}), this then risks the continued propagation of non-consensual imagery. The above highlights the need to focus beyond just the models. Much of the critical AI literature focuses on vital interventions to change model outputs that discriminate against protected classes. This work, instead, more closely follows work such as Buolamwini \& Gebru’s "Gender Shades"~\cite{buolamwiniGenderShadesIntersectional2018} that is focused on the \textit{data} which is fundamental to AI system development.

To understand the risks posed by using synthetic data to create and add to datasets used in machine learning development, it is necessary to understand the landscape of both machine learning development and the stakeholders impacted by it. We will use the taxonomy of dataset development stages and subjects presented by Khan \& Hanna~\cite{khanSubjectsStagesAI2022a}. We lean on this taxonomy throughout the paper, finding it to be clear in its intention of “providing a common language for conversations across datasets” between practitioners, scholars and regulators. Starting from first principles, machine learning is not rules-based like traditional software development, but instead consists of a practitioner teaching a model to identify patterns in a dataset. To begin, Khan \& Hanna assert, the \textit{task} for which the model is being trained must be formulated and constrained. Next, \textit{data} must be collected, meeting the constraints of what is necessary to train a model to achieve said task. That data collection is usually broad, requiring the data be \textit{cleaned} before it is annotated. The \textit{labels} attributed to data by the annotator are of vital importance to machine learning systems, as the systems are taught to identify those labels in their training data. After this point, model \textit{training, valuation, implementation}, etc. may occur. Khan \& Hannah define multiple \textit{stakeholders} in the process of creating the datasets used for ML development: the curator, the data annotator, the data subject, the copyright holder, and the model subject. The \textit{curator} is the entity responsible for dataset creation, while the \textit{annotator} is (frequently outsourced~\cite{grayGhostWorkHow2019}) responsible for annotating the dataset. The \textit{data subject} is the person whose biometric information is present within the collected data, the \textit{copyright holder} may hold exclusive rights over that data\footnote{not all data is copyright protected, and even when it is, different legal regimes have different limitations and exceptions to exclusive intellectual property rights}, and the \textit{model subject} is the person who is impacted by the decisions made by the model trained on the data. The last three categories are fluid, and can consist of the same person or of two or three distinct entities. 

Creating a training dataset that is representative of model subjects is challenging, and when done poorly, results in inaccurate and frequently harmful outputs; here we consider how such harm might arise across different stages. In producing a representative dataset, different dimensions of identities are frequently missed. Datasets are frequently biased by necessity to meet a specific intended use of a model, but when categories are socially constructed (such as race and gender), the observed, inferred data that dataset curators use to bound and constrain data collection (and that annotators must use to label) can clash with how the data subject self-identifies. Biased data representation is also a concern, with annotation reflecting social biases and stereotypes across gender, race, and more~\cite{scheuermanHowWeVe2020}. This can lead to rampant misrepresentation and miscategorization of both data and model subjects, producing forms of control. Annotation work is inherently an interpretive project, but results in data that is perceived to be ground truth~\cite{bowmanWhatWillIt2021}. In reality, as shown by Recht et al. in work on testing the generalizability of ImageNet, when attempts to replicate annotation are made, different distributional properties for the same data emerge~\cite{rechtImageNetClassifiersGeneralize2019}. Annotation is also frequently the cause of artifacts in datasets that allow for models to overfit to training data when solving a task, with many of the concerns arising from how human data annotators are instructed to label~\cite{yangFairerDatasetsFiltering2020}. Beyond concerns with annotation, the question of what data to collect when constructing a dataset to correctly answer a question is complex. The task of annotation itself presupposes that the question being asked of a model is one that can be answered --- Aguera y Arcas et al. demonstrate how a model trained on ‘gaydar’ data was in reality labeled around stereotyped aesthetic traits, showcasing an example of labels being generated not because of any model-pertinent aspects of the data, but rather simply because the question was being asked~\cite{arcasAlgorithmsRevealSexual2018}.

Beyond the above challenges to creating a representative dataset, there are ethical issues that are raised by efforts to produce a dataset that represents a diverse community. Data collection requires infrastructure, and that infrastructure is frequently co-constitutive with surveillance infrastructure. Even when data collection is initiated in service of providing services to the most disenfranchised, rendering the members of those communities hyper-visible frequently serves to hurt those same communities, as decisions are made for them by others~\cite{andrusDemographicReliantAlgorithmicFairness2022}. These decisions can reinforce oppressive norms, such as visual gender binaries~\cite{bivensGenderBinaryWill2017,hamidiGenderRecognitionGender2018}, further delegitimizing disenfranchised groups in a clear example of administrative violence~\cite{spadeNormalLifeAdministrative2015a}. Even when categorization schema of data subjects are correct, their use as prescriptive instead of explanatory can lead to attribution errors, co-opting classification in a oppressed group as a reason for that very oppression. Machine learning systems used to predict recidivism are a prime example~\cite{christinAlgorithmsPracticeComparing2017}, where factors like race, which make a group member more likely to be targeted for discrimination, are frequently used instead as a predictive factor when individuals are made model subjects. 

\subsection{Participation, Consent and Privacy}~\label{RW:PCR}
Participatory approaches are frequently fronted as a way of mitigating the harm that results from AI systems, both at the dataset level as described above, and in model training and deployment. These approaches focus on engaging the public, and build on policy approaches such as feedback sessions, public hearings and impact assessments~\cite{gluckerPublicParticipationEnvironmental2013,hugelPublicParticipationEngagement2020}. Participatory design methods in particular focus on co-design to incorporate user context, needs and values~\cite{bratteteigUnpackingNotionParticipation2016, sloaneMakeAIFair2022,iversenValuesledParticipatoryDesign2012,halloranValueValuesResourcing2009}, designing systems \textit{with} those affected instead of \textit{for} them. Recently, participatory AI work has explicitly focused on those for which AI most frequently exacerbates harm~\cite{irgensDesigningYouthParticipatory2022,robertsonWhatIfDon2020a}. Patel at al.~\cite{patelParticipatoryDataStewardship2021} draw from previous work, including Arnstein’s influential Ladder of Citizen Participation~\cite{arnsteinLadderCitizenParticipation1969}, to detail five levels of participation in data stewardship, including: 1) \textit{informing} people about how their data is used through methods such as model cards, 2) \textit{consulting} people through UX research and surveys, 3) \textit{involving} people in data governance through panels and public deliberation, 4) \textit{co-design} of data governance and consequent technologies through structures such as data trusts, and 5) enabling \textit{decision-making} through citizen-led governance boards. We will return to Patel's framework when discussing the risk of synthetic data circumventing consent (Section~\ref{R2}). 

There is a wide --- and growing! --- diversity of participatory work in AI. Examples range from crowdsourcing impacts~\cite{barnettCrowdsourcingImpactsExploring2022,diazCrowdWorkSheetsAccountingIndividual2022} and data labeling~\cite{parkAIBasedRequestAugmentation2019} to eliciting preferences for dataset collecting and design decisions~\cite{christianoDeepReinforcementLearning2023}. Peng et al.~\cite{pengMitigatingDatasetHarms2021} recommend that dataset creators make ethically salient information clear and accessible while actively stewarding the dataset and its future use, and encourage retrospective study of datasets due to the difficulties in understanding issues at the beginning. Hutchinson et al.~\cite{hutchinsonAccountabilityMachineLearning2021} detail documentation requirements at each stage of the dataset development lifecycle, with different document types for each stage, and call for frameworks for transparency and accountability. These fall across the range of Patel et al.’s framework,~\cite{patelParticipatoryDataStewardship2021} and critiques of these methods include characterization of it as ‘participation-washing’~\cite{gilmanWindowDressingPublic2022,sloaneParticipationNotDesign2020}, with Arnstein describing approaches such as public requests for comment as “tokenizing” and “inadequate in shifting power”~\cite{arnsteinLadderCitizenParticipation1969}. Sloane et al.~\cite{sloaneParticipationNotDesign2020} argue that these approaches can function as unrecognized labor, and the line between tokenization and participation in cases such as crowdsourcing is quite blurry. Birhane et al.~\cite{birhanePowerPeopleOpportunities2022a} show examples of community inclusion in annotating datasets, improving documentation and increasing the utility of large language models for under-served languages, and other examples include community organizations such as the Detroit URC3 which evaluates potential partnerships between community organizations and researchers to avoid exploitation~\cite{corbettPowerPublicParticipation2023}, and examples from Indigenous Data Sovereignty~\cite{rainieIndigenousDataSovereignty2019}. At a large scale however, there are still major hurdles. Groves et al. investigate the hurdle of making participatory approaches work in the commercial AI labs that are the primary site for AI research, and find that “corporate profit motive and concern around exploitation are at present functioning as significant barriers to the use of participatory methods in AI”~\cite[p.~10]{grovesGoingPublicRole2023}.

While the above participatory approaches center shifting decision-making power to include data subjects and model subjects, these approaches tend to require model creators to opt in at least for now. Consent, though enacting a more limited form of participation, \textit{requires} model creators to be wary of unfair and deceptive practices that overstep expressly-informed consent when collecting and using data. Indeed, consent violation is a legally cognizable privacy harm, one with potential repercussions. In the U.S., state information privacy laws do some work to enforce this, with the Illinois Biometric Privacy Act (BIPA) both resulting in a significant number of lawsuits alleging violation, and being responsible for the largest settlement amounts from companies who have breached BIPA by deploying FRT~\cite{stricklerRecentDevelopmentsPrivacy2020,yewRegulatingFacialProcessing2022}). For instance, in \textit{Vance v. IBM}, the court affirmed that IBM violated BIPA by not receiving written consent before collecting and disseminating individuals' images in their "Diversity in Faces" dataset~\cite{goldenfeinPrivacyLooseGrip2023}, even though the images used were public. Publicly accessible personal information comes with an intended context of use, which can be violated by memorized and regurgitated data~\cite{carliniExtractingTrainingData2023}. As will be discussed in Section~\ref{R2:FTC}, to date, the Federal Trade Commission’s enforcement power around unfair and deceptive data practices has centered upon the absence of consent. Practically, this instills consent as the most direct way for data subjects and model subjects to participate in decision-making around the models which affect them, albeit mostly \textit{ex post facto} through their ability to prompt enforcement when discovering that their consent has been violated. 

Consent violations frequently occur through improperly scoped consent, where data collected for one purpose is repurposed. This can result in adverse effects beyond the concrete privacy harms~\cite{solovePrivacySelfManagementConsent2012} that are most often legally enforced in cases such as data breach, e.g., identity theft. As an example, data used beyond its consented purpose leaves data subjects at risk of discrimination harms, facing miscategorization and expansion of surveillance, as detailed above. Such data can also be sold and shared with third parties, further increasing the odds that it is not being used for the purpose it was collected, and therefore that it is frequently in violation of the consent of data subjects~\cite{caloBoundariesPrivacyHarm2011}. Even when consent is nominally obtained, transparency is often in name only, with data subjects overwhelmed by opaque and all-encompassing digital policies, terms, and conditions~\cite{pasqualeLicensureDataGovernance2021}. 

Ultimately, the question of consent is complex. Brown et al.~\cite{brownWhatDoesIt2022} argue that the current paradigm of training on publicly accessible data makes it highly challenging to distinguish what public data was made public with blanket versus contextual consent, and that, therefore, obtaining informed consent is difficult at best. They make the case for training solely on data explicitly consented for public dissemination. We will return to this argument in Section~\ref{Dis}, as it supports using synthetic data generated from properly-consented real data or responsibly procedurally created data, and criticizes using synthetic data generated and used in a manner that exacerbates concerns around consent.

\subsection{Synthetic Data}~\label{RW:SD}
Synthetic data in machine learning is defined by its driving goal of mimicking real-world data --- it is synthesized to be used as though it were real data for training machine learning algorithms~\cite{jacobsenMachineLearningPolitics2023a}. It differs from what is usually referred to as ‘data’, i.e. non-synthetic data, in that it does not have an explicit 1:1 real-world referent. When training a computer vision model to recognize a face, the data traditionally used are representations of real faces, photographs taken of real people. The same holds true for other forms of data --- scientific data records representations of physical phenomena such as sensor readings, natural language data is text composed by a real person, etc. Synthetic data is made to resemble these things, but is not explicitly a representation of a real thing. Using the example of a face, a synthetic face could be a drawing that looks for all intents and purposes like a face, but that is does not represent a specific person.

In actually creating synthetic data, however, things become muddier. The term encompasses data generated by generative models, more traditionally augmented data, and procedurally created data~\cite{liuDeepLearningProcedural2021}. We differentiate between the first two and the latter category based on how derivative of a real-world training dataset they are. \textit{Generated data} is the output of generative models: ML systems that produce a (supposedly novel ~\cite{carliniExtractingTrainingData2023}) output from an input by abstracting over their training data. One generative model that has captured popular attention~\cite{perezChatGPTLearningTool2023} is StableDiffusion, which generates art from a user-provided input sentence~\cite{caoSurveyGenerativeDiffusion2023}. \textit{Augmented data} is fuzzier, but equally derivative of a real-world training dataset; the term tends to refer to any real-world data to which modifications have been made. A model creator seeking to increase the performance of a model on its specific task may create many versions of each image sampled from the input dataset, creating augmented data. This type of synthetic data cannot be considered inherently private or unbiased, with generative models explicitly being found to frequently regurgitate memorized training data~\cite{baiTrainingSampleMemorization2021}. At larger scales, this type of data can be used as training and evaluation datasets too. As detailed by by Khan \& Hanna~\cite{khanSubjectsStagesAI2022a}, datasets are vital components the larger model development cycle, priming synthetic data to reinforce and scale skewed values and requirements that are embedded within models, datasets, and benchmarks.

Generated and augmented data differs from techniques for \textit{procedural creation} of data, where dataset designers make active decisions to create ‘net-new’ representations of data similar to what might be found in the natural world. There are some important differences. The easiest way to visualize how procedural creation works is considering video game character creation --- a player starts with a base face shape, and adds the features they want. Notably, with procedural creation of faces, the base volumetric face scans that are used are very far removed from the people that were scanned --- they are low fidelity representations, making data lineage even murkier. Instead of outputs being generated from, and therefore bound by training data, procedural creation can result in ‘net-new’ data that never existed previously.
Or, as an example outside of computer vision and facial recognition, consider procedurally created finance data which uses agent-based models to mimic real world data generation by creating representative agents and attempting to model money laundering~\cite{lopez-rojasMoneyLaunderingDetection2012}. It must be remembered, however, that both agent- and procedural-based synthetic data are highly determined by preconfiguration and design. As such, making inferences about the real world based on procedurally created data is difficult at best. Recalling Jordan et al.'s framework of synthetic's data usability, this type of synthetic data faces utility and fidelity hurdles ~\cite{jordonSyntheticDataWhat2022}.

Synthetic data has frequently been explored as a method to avoid privacy concerns, increase model performance and to reduce model bias. Privacy concerns have the longest history of motivating synthetic data~\cite{jordonSyntheticDataWhat2022}. Healthcare~\cite{gonzalesSyntheticDataHealth2023} and financial~\cite{assefaGeneratingSyntheticData2021} domains have been particularly attracted to synthetic solutions due to the sensitivity of their data. Examples including simulation studies in population health~\cite{nguforMixedEffectMachine2019}; synthetic clinical records used for IT development, education, and training~\cite{davisUsingMicrosimulationCreate2010}; money-laundering detection~\cite{lopez-rojasMoneyLaunderingDetection2012}; and public release of augmented financial and healthcare datasets to enable open science and research~\cite{harronLinkingDataMothers2016}. In contexts of societal bias, synthetic data has been explored as a way to remove disparate impact~\cite{feldmanCertifyingRemovingDisparate2015,kamiranClassifyingDiscriminating2009,zhangCausalFrameworkDiscovering2016}, to suppress imbalance effects and to racially balance datasets~\cite{kortylewskiAnalyzingReducingDamage2019}, as well as to remove sensitive information and blind models to race~\cite{wangApproachingMachineLearning2019}. However, the latter has been found to not always be effective in practice, with applying a ‘veil of ignorance’ not having any notable influence on accuracy of FRT on under-represented categories~\cite{wehrliBiasAwarenessIgnorance2022}. 

Increasing model performance by using synthetic data has usually meant enlarging datasets to provide robustness to outliers~\cite{wongUnderstandingDataAugmentation2016,fawazDataAugmentationUsing2018,daiGoodSemisupervisedLearning2017}. Additionally, operating at a slightly different scale of enlargement (from 0), it has been used in situations where real world data is difficult to access. Google recently demonstrated AlphaGeometry, an AI system purported to solve “Olympiad geometry problems at a level approaching a human-gold medalist”, trained solely on a dataset of 100 million synthetic math proofs --- a dataset which could not exist using human generated proofs~\cite{trinhSolvingOlympiadGeometry2024}. Computer vision systems require (often impossibly) large amounts of labeled training data in a specific domain~\cite{birhaneLargeImageDatasets2021}. 

Finally, consider the context of facial recognition. Though large, open datasets specifically developed for facial recognition tasks exist, such datasets are either extremely basic (i.e. passport photos with great lighting), too narrow (only contain a biased subset of race, gender, head/body pose, etc.), or simply contain glaring and challenging shortcomings~\cite{rajiFaceSurveyFacial2021}. As a result, dataset creators utilize synthetic data. In FRT, this is either (1) procedurally created ‘synthetic’ data, where a bone structure scan is used as a basis for volumetric face models, and then textures representing features are stretched across that model and swapped out~\cite{yiLearningFaceRepresentation2014}, or (2) the perturbation of existing data to produce more diverse samples from the existing distribution, including both simple techniques and advanced techniques such as diffusion models and generative adversarial networks (GANs)~\cite{dhariwalDiffusionModelsBeat2021}. We expand upon the use of synthetic data for FRT in the following section.

\section{Risk 1: Diversity-Washing}~\label{R1}
This section presents an example of using synthetic data; first describing a dataset that is partially-synthetic, a blend of augmented data and real data, and then describing a dataset that is procedurally created synthetic data. In both cases, we detail how using synthetic data risks creating datasets --- and subsequently training and evaluating models from that data --- that fail to mitigate bias in data distribution and representation. Furthermore, there is a risk of propagating harm through a patina of legitimacy, and through diversity-washing potentially harmful models.

The example we use is a real world example, where one of the authors had previously created synthetic datasets to evaluate facial recognition technology (FRT)~\cite{normanEvaluationForensicFacial2023}. We present this example to illustrate a risk of synthetic data and ground it in a real world setting --- this is not an attempt to present novel work on FRT evaluation. 

To provide a brief background: FRT is created with the aim of matching images of identifying faces. Companies that sell facial recognition such as Clearview often tout accuracy rates of their systems of 97\% or more~\cite{tabohFacialRecognitionTechnology2021} --- but these calculations are often made under ideal conditions. In real world conditions, such as surveillance camera footage, captured images of faces may be poor quality. FRT has been shown to be prone to make erroneous matches (i.e. identifying someone incorrectly as a match) when using low quality images as input~\cite{huInfluenceImageQuality2021}. However, users of these systems, such as the police, and adjudicators such as judges or members of Congress, who are not experts in FRT or ML/AI, are at a distinct disadvantage in evaluating companies' claims.

\subsection{Augmented Partially-Synthetic Dataset}~\label{R1:A}
In order to evaluate FRT models in real world settings, first, benchmark performance for FRT models needed to be established. This occurred by stimulating the (highly unreliable) process of a human identifying an individual from a visual lineup of other humans with similar characteristics. To do so, a source image of a selected identity was identified from the base dataset, detailed below, and a “digital lineup” of (mathematically) similar faces from that base dataset were created. Augmented data was then created by progressively degrading the image of the source identity, and then this augmented data was compared to the similar identities in the digital lineup, as well as to the source, in order to mimic real world settings. The success of the evaluated models was defined by the rate the correct identity was selected with the augmented data, the degraded source images, as input. 

A significant body of knowledge already exists concerning both the obvious and non-obvious potential harms in gathering image data containing human subjects, and the real harms of processing such information through FRT~\cite{rajiFaceSurveyFacial2021}. As such, it was important to begin with core datasets that had already been evaluated thoroughly in the literature, rather than collect wholly new human subject data. As such, CASIA-Webface, one of the two most widely used and evaluated public datasets~\cite{yiLearningFaceRepresentation2014,kawulokAdvancesFaceDetection2016}, was selected as the dataset for use as both non-augmented data and as the base for augmented (in this case, degraded) data. This was chosen due to its sourcing from crawled and scraped publicly available images of celebrities, strict rules prohibiting commercial use, wide variation in image quality, large number of identities (depth), and large number of images per identity (width) --- important features for the dataset that was the provenance for later augmentation. As detailed in Section~\ref{RW:DDR}, this choice of base dataset is inherently political despite frequently being rendered neutral. Using it as the base for the creation of synthetic data makes it inherently more so due to the downstream effects of using the base dataset in generating derivative data.

The data augmentation techniques used to generate the augmented portion of this mixed dataset can differentially and unpredictably scale issues --- making images black and white, as an example, could further segment training data by racial presentation. Model creators frequently augment data while training models, past the stages of development where they are considering dataset collection and annotation. This process has a set of well documented risks for model fairness. However, in using this process to create large base datasets, there is a change of framing that re-introduces these risks. For example, the dataset created for the FRT evaluation (created by augmenting data and combining with real data) was created with the explicit goal of being more representative of real world conditions. Datasets are frequently treated as ground truth~\cite{birhaneLargeImageDatasets2021}, hiding the decisions and processes by which they were created. This risks ignoring issues that can occur from augmenting data. Even if synthetic data appears 'diverse', the generation of that data cannot be unwound from the particular datasets and models that it is being generated from, and any attendant shortcomings. To start, any biased representations would at best replicate from the original dataset. If data augmentation technique(s) impacted some subjects differently than others, the resulting impact could be unintended bias in the dataset. Since the presence of such relationships are rarely known, much less understood statistically in datasets, it is also possible that the sampling strategy used to choose which data from the dataset will be used in training may actually exacerbate harm by over-representing biased representations. Deep learning techniques are generally already susceptible to overfitting, where a model learns how to predict patterns in a way that pays too much attention to the training data, and doesn't generalize to other data --- it learns specific idiosyncrasies and meaningless data artifacts. Synthetic data seems like it should have the capacity to remedy overfitting, through careful and bespoke dataset construction that debiases data distributions. One could assume this would increase fidelity and enable better generalization over a more diverse training space. In reality however, when synthetic data is overfitted, these idiosyncrasies can go through the entire model training process unnoticed. As such, synthetic data instead increases the likelihood of overfitting errors being propagated through, necessitating that further technical care is taken to prevent overfitting. In our FRT evaluation example, such preventative measures were taken by curating the dataset so that visible artifacts such as skin tone were at parity with acceptable real world dataset distributions. However, the perils of overfitting are a way in which synthetic data can struggle to meet the standard of utility necessary to work as a replacement for real world data~\cite{jordonSyntheticDataWhat2022}.

\subsection{Procedurally Created Fully Synthetic Dataset}~\label{R1:P}
Beyond the above partially-synthetic dataset, there was a need to better evaluate performance on specific types of data not present in the original datasets that the FRT models were trained on. So, a synthetic dataset consisting of procedurally created data, namely mixed examples of non-degraded and degraded computer-generated faces~\cite{qiuSynFaceFaceRecognition2021,basakLearning3DHead2021}, was developed. As previously detailed, this is a common usage of synthetic data --- needed representative data was not available for collection, and so generating synthetic data was the easiest method of proceeding.

The Synthesis.AI software\footnote{Synthesis AI (https://synthesis.ai)} used to create the fully synthetic dataset (as with most procedural synthetic human/object generation tools) works by providing unprecedented control over how a dataset and its inherent metadata parameters are specified. This software employs a combination of classic rendering and generative synthesis to create photo-realistic images of human faces, bodies, and environments~\cite{nikolenkoSynthesisHumansCreate2022}. A user is able to decide how much and which type of each characteristic (in our case age, race, gender, hair type, pose, lighting etc.). However, the tool did not make any suggestions or place any controls based on sociotechnical norms or demographic data (such as the census etc.) when creating a synthetic human dataset of any type. When first testing the Synthesis.AI API, a dramatically racially imbalanced dataset was returned, even though the specification given was for randomization of the race characteristic. At first glance, the dataset appeared diverse and was numerically at parity for gender. However, the software lacked permutations for Asian people, Middle Eastern people and Black women, leading to a stark racial disparity upon deeper inspection, and a preponderance of white men and white women despite attempts at balancing racial demographics. Such a system allows any user to easily create an unintentionally biased dataset, which could then be used to train a biased model. Instead of mitigating data distribution and representation concerns, this risks extending them.

As a further example, Microsoft’s FaceSynthetics~\cite{woodFakeItTill2021} is a procedurally created synthetic dataset of 100,000 individuals, with faces derived from representative 511 base scans. However, these 511 base scans include only ~30 Black men, and even fewer Hispanic/Arab/Indian men and Black/Hispanic/Arab/Indian women (borrowing the reported demographic categories), meaning that the fully diverse population they claim include multiple racial categories fully defined by the ways in which these <30 faces can be manipulated through a generative process. These manipulations include fine-tuning hair, expression, and clothing, but published details on the process of how these potentially racially-coded aspects were chosen are sparse. It is not known how those features are distributed in real faces, and attempting to extrapolate a portion of a representatively diverse dataset from such a small set of base faces leads to a risk of \textit{statistical diversity without representational diversity}, compared to a representative dataset of real images with both statistical and representational diversity. Synthetic data here falls flat in addressing these complex, cultural and deeply contextualized factors.

These tools risk falling into the ‘panacea of legitimization’ that Frank Pasquale describes~\cite{pasqualeLicensureDataGovernance2021}, where ethical concerns are not only routed around, but are routed around in such a manner that they can reify malpractice due to the co-constitutive nature of ML practices and computing platforms~\cite{bermanMachineLearningPractices2023}. We point to recent work focused on toolkits for supporting practitioners in contextualizing ML system work as an avenue for improving upon this~\cite{dengExploringHowMachine2022}.

\section{Risk 2: Circumvented Consent}~\label{R2}
Consent has become a key component of privacy and data protection, both via regulatory enforcement and as a necessary foundation for the participatory approaches that have emerged as ethical practice for preventing harm. Consent also plays a role in U.S. sectoral statutes such as HIPPA, U.S. state laws such as California’s California Privacy Rights Act and Illinois’ Biometric Information Privacy Act, as well as laws in many other countries, with the EU’s GDPR a notable example. These statutes share a common goal: to provide people with control over their personal data, via notification, access, and consent regarding the collection, use, and disclosure of personal data. This type of privacy regulation is referred to as “privacy self-management” by Solove~\cite{solovePrivacySelfManagementConsent2012}, and focuses solely on whether or not data subjects have consented, rather than on value judgements of privacy practices. This section will focus on illustrating the risk that synthetic data poses to consent-based frameworks by expanding upon how the Federal Trade Commission (FTC) has functionally used consent as a key aspect of conducting enforcement against companies using ML systems. The analysis is guided by one author’s experience at the FTC, but draws upon solely public knowledge.

~\subsection{Consent, Deception and Model Deletion}~\label{R2:FTC}
The FTC plays a vital role in the current U.S. privacy legal framework. This framework emphasizes individuals' notice of, and consent to, the collection and use of their data. The FTC is an independent agency of the United States government that is tasked with protecting consumers and promoting competition in the marketplace. In the absence of federal privacy law, the FTC has played the role of de facto privacy enforcement, primarily based on its authority to police unfair and deceptive business practices. The Federal Trade Commission Act, and specifically Section 5, is a broadly applicable federal statute prohibiting “unfair or deceptive acts and practices”~\footnote{15 U.S.C. Sec. 45(a)}. An \textit{unfair} practice ”causes or is likely to cause substantial injury to consumers which is not reasonably avoidable by consumers themselves and not outweighed by countervailing benefits to consumers or to competition”, while a \textit{deceptive} practice includes “any ‘representation, omission, or practice’ that is (i) material, and (ii) likely to mislead consumers who are acting reasonably under the circumstances”~\cite{soloveFTCNewCommon2014}. Notably, deception does not require any proof of intent. The FTC has brought deception claims against companies who have violated the terms of their privacy policies, failed to uphold promises of data security, or have failed to provide sufficient notice regarding data collection and use~\cite{soloveFTCNewCommon2014}.

In settling cases against companies that have deceptively collected data, the FTC has required not only that the data in question be deleted and the affected users be notified, but also that all "affected work product"~\cite{liAlgorithmicDestruction2022} be deleted as well --- including models trained on that data. This approach is referred to as \textit{model deletion}. The FTC posits that this approach is necessary in order to ensure that companies do not profit from the unfair or deceptive collection of data, and to prevent them from using the data in the future. Intellectual property is typically a tech company’s most valuable asset; it is an important factor for securing venture capital funding and the sale or licensing of IP often comprises tech companies’ core business models. In forcing a company to delete models, the FTC has also significantly changed the deterrence calculus for companies: from paying (relatively) small fines, to potentially losing a vital business asset~\cite{elderWrongfulImproversGuiding2022}. The FTC has used the concept of model deletion in recent enforcement actions against companies that have collected data deceptively, including Amazon Ring, RiteAid, Everalbum, Clearview and Kurbo (WeightWatchers) ~\cite{hutsonAmericaNextStop2022}. 

\subsection{Synthetic Data and FTC Enforcement}~\label{R2:D}
The use of synthetic data risks undermining the utility of deception-based enforcement in regulating the collection of data, and therefore also undermines the regulation of models trained on such synthetic data. As previously described in Section~\ref{RW:DDR}, datasets play a foundational role in the models trained on them, and trusting models trained on deceptively collected datasets to operate without harm seems foolhardy. Enforcement by the FTC has hinged upon arguments that data was collected and used deceptively --- often argued due to the absence of proper consent. By using synthetic data, however, it becomes easy for model creators to obfuscate the origins and consent of the data being used to create models. In the case of a procedurally created synthetic dataset, consent is no longer a procedural hook to limit downstream harms flowing from use, while in other synthetic datasets, unless data lineage is carefully recorded, traceability to the original data is at risk~\cite{scheuermanHumanDataDataset2023}.

Synthetic data also exacerbates existing logistical challenges for model deletion as an enforcement tool. Synthetic data brings questions of data lineage to the forefront, as ever-more-complicated sets of original, augmented, and derivative data are produced based on new face datasets with millions of people. As a small example, keeping track of whether a single version of a dataset has undergone ethical testing, or was sectioned off as a test dataset, is a challenge for FTC enforcers --- let alone when datasets include different scaling factors and different degradations, with different subsets of identities (generated, procedurally created or real) and different levels of augmentation. One of the key hurdles to model deletion is the requirement for a high level of internal company documentation and logging. This documentation and logging is essential to identify the data that was collected illegally or deceptively, as well as the work product that was developed using that data. However, not all companies have robust internal documentation and logging systems, which can make it difficult for the FTC to determine the extent of the harm caused by the illegal or deceptive data collection practices. Another challenge is authentication and audit. Companies must demonstrate that they have successfully deleted the affected work product, and the FTC or other enforcement bodies must have a way of verifying that the company is being honest. However, this can be difficult, as it requires a level of trust in the company and its processes. In a setting where the FTC has to this point relied on settlement agreements, synthetic data further complicates existing logistical challenges, presenting important friction to enforcement.

~\subsection{Beyond Deception}~\label{R2:BD}
In considering enforcement against companies using ML systems, it is important to note that beyond deception, the FTC has also enforced its unfairness authority. This occurred in the case of RiteAid, where a biased facial recognition model was used to falsely identify people in certain protected classes as more likely to commit crime. This case was first-in-kind, but demonstrates that the FTC is not solely beholden to deceptive data collection as an avenue for enforcement. However, the details of the case were particularly egregious, with RiteAid failing to undertake even the most basic risk assessments, and in part hinged on a violation of a previous settlement. Additionally, the system was trained on in-store camera footage without consent from data subjects, and model subjects were not notified or able to opt-out. Thus if RiteAid had been investigated for deceptive data collection, harm resulting from this system could have been prevented at the point where the system was trained non-consensually. But what would happen if RiteAid had trained its FRT model on synthetic data? FTC enforcement would have to hinge on unfair practice alone. While successful here, the case against RiteAid was egregious. The Supreme Court’s neutering of the FTC’s power to levy fines~\cite{chopraCaseResurrectingFTC2021}, in addition to both the deception and unfairness enforcements occurring through settlement rather than being decided in court, means the boundaries of the FTC’s ability to engage in this type of enforcement are still undefined. As such, the FTC's ability to intervene both at the data collection stage (deception) and the model deployment stage (unfairness) gives options\footnote{While there exists the potential for both unfairness at data collection and deception at model deployment, cases to-date have lined up in this fashion}. Synthetic data complicates the usage of a demonstrably useful tool for protecting data subjects and model subjects, by complicating the use of the deception standard.

Finally, it cannot be forgotten that while risking increased friction and obfuscation, synthetic datasets composed of augmented or generated data are demonstrably derivative, inherently based on real data representing real data subjects. And, despite the veneer provided by language such as ‘net-new’, procedural creation of synthetic data is also derivative, and thus suffers from issues of consent and participation too. In the example discussed in Section~\ref{R1:P}, the procedurally created synthetic dataset for FRT evaluation, the dataset was generated using Synthesis.AI's commercial software. Software tools such as Synthesis.AI often utilize face and body scanning technology as the foundation of their generative processes, raising concerns around the limits of informed consent. The data subjects upon which these technologies are trained are rendered invisible and thus the use of such software is predicated on, at best~\cite{rajiSavingFaceInvestigating2020}, ambiguous consent. Similarly, it is important to acknowledge that the data subjects whose likenesses are captured in the CASIA-Webface, while primarily scraped from public sources, were not asked for informed consent regarding their data's use --- even when it has been shown that having your face included in such a dataset increases the accuracy of facial recognition models on your specific face~\cite{dulhantyInvestigatingImpactInclusion2020}. As detailed by Peng et al.~\cite{pengMitigatingDatasetHarms2021}, using derivatives of common datasets introduces scaling concerns around propagation of improperly consented data, and as such, using synthetic data risks \textit{further} scaling propagation of this issue. In decoupling data subjects from their data, this also removes their capacity to participate. Reconsider the Participatory Data Stewardship Framework mentioned above in Section~\ref{RW:PCR}; all five levels require that data subjects have at the very least visibility, and preferably control, over their data ~\cite{patelParticipatoryDataStewardship2021}. In further removing the ability for data subjects to consent, not only is that minimal level of agency reduced, but the potential for involvement in decision-making that directly effects them is erased. 

\section{Discussion}~\label{Dis}
The positioning of synthetic data as a panacea to problems of representation and deceptive data collection, furthered by its portrayal as synthetic, as neutral, as created without lineage, risks placing the means of fixing those problems in the hands of those who created them and trusting that they'll get it right. Instead, the kinds of racialized misrecognition and bias that Ruha Benjamin, Safiya Noble, and others have drawn attention to must be considered when determining whether to use synthetic data. As Benjamin argues, our current machine learning development ecosystem must reckon with a history of discriminatory design in which racist values and assumptions are built into our technical systems. The 'new Jim Code', as she terms it, works to deepen the production of disparate harm, even while cloaked in neutrality and the language of innovation~\cite{benjaminRaceTechnologyAbolitionist2019a,iraniChasingInnovationMaking2019}. Discriminatory practices are inherent to the current state of AI system development, privileging whiteness and discriminating against people of color, specifically women of color~\cite{nobleAlgorithmsOppressionHow2021}. Sara Ahmed’s work on the phenomenology of whiteness highlights the danger of a solution that further enables a reification of non-whiteness as a space outside. She details that "institutional spaces are shaped by the proximity of some bodies and not others: white bodies gather, and cohere to form the edges of such spaces”~\cite{ahmedPhenomenologyWhiteness2007}. Synthetic data as a fix in this racialized context risks further enabling amplification of racial hierarchies, allowing for those within the boundaries to actively constitute the exclusionary and weaponized edges of these spaces: to define a face, train a model based on that definition, and decide its performance based on labeling racial boundaries. It risks not alleviating but instead contributing to
\textit{race as a technology,} designed to “stratify and sanctify” social injustice in the architecture of everyday life~\cite{benjaminRaceTechnologyAbolitionist2019a}; an added consolidation of power. 

Another example of the risks of consolidation of power through synthetic data arises when considering the inherent relational aspects of data privacy. Solon Barocas and Helen Nissenbaum identify the risk of a "tyranny of the minority" in big-data analytics when "the volunteered information of the few can unlock the same information about the many"~\cite{barocasBigDataEnd2014}. More recently, Salome Viljoen emphasizes the importance of a relational theory of data governance~\cite{viljoenRelationalTheoryData2021}. As Viljoen explains, dataflows entail not only ‘vertical’ relations between a particular individual and a data collector, but also ‘horizontal’ relations between the individual and others sharing relevant population features. Viljoen focuses on the manner in which informational infrastructures rely on group classification to make sense of individuals by taking a ‘relevant shared feature,’ generating a prediction based upon that shared feature, and then applying this prediction. When those shared features are derived from synthetic data, decoupled from any real context and perhaps even specifically created to rectify gaps in representation, we hand power to those creating that synthetic data. We risk imposing designers' decision-making in lieu of and upon those least likely to have been represented and most likely to be harmed by both the diversity-washing and the side-stepping of consent. After all, if they were represented or able to consent in the first place, there would be no need for additional synthetic data. Data minimization and lineage principles~\cite{hutsonAmericaNextStop2022} are a first step towards mitigating this issue by requiring documentation and its requisite transparency into where data has come from. The need for this is also readily apparent when considering contexts such as the EU's Right to be Forgotten~\cite{razmetaevaRightBeForgotten2020}, where synthetic data further complicates the ability to be removed from a dataset. In making it harder to decouple data from its context through the use of synthetic data, there is an avenue for mitigating consolidation of power and ensuring consent. 

Many of the risks discussed in this paper propagate from the centralized decision-making nature that synthetic data imposes. Participatory governance structures, as mentioned in Section~\ref{RW}, offer a potential solution here --- synthetic data could be created to represent concerned groups \textit{by} those self-same groups, re-establishing control and mitigating some concerns around consent and contextualization. Many have called for training data to be restricted to only data that is explicitly consented to be used, though consent is difficult (if not impossible) to establish and propagate over multiple degrees of separation~\cite{brownWhatDoesIt2022}. But synthetic data, when generated to purpose by concerned communities, can provide a potential solution. Models such as those presented in the field of Indigenous Data Sovereignty, where there has been effective push back against external categorization schemas~\cite{rainieIndigenousDataSovereignty2019} show potential for participatory governance models to address group misrepresentation~\cite{andrusDemographicReliantAlgorithmicFairness2022}. 

Additionally, there are practical considerations that make 'participatory synthetic data' an attractive path forward. Both language and computer vision models are beginning to contend with a phenomena commonly referred to as ‘garbage-in garbage-out'~\cite{shumailovCurseRecursionTraining2023,martinezUnderstandingInterplayGenerative2023}. This refers to the advent of generated data becoming commonplace and public, and the related struggles by those capturing data to differentiate between that data and real data, leading to data generated by a system such as ChatGPT becoming its own training data in the future. Work by Agnew et al.~\cite{agnewIllusionArtificialInclusion2024} examines the use of these models to replace participants in industry research, highlighting how proposals to do so are motivated by cost reduction and data diversity. They identify these proposals as facing issues in aligning with the values human participants identify as important, specifically including inclusion and representation, necessitating further contextualization and bespoke dataset creation. In such a world, large tech companies may have business motivations for engaging with responsibly created synthetic data, and as demonstrated by Deng et al.~\cite{dengExploringHowMachine2022}, methods exist for enabling machine learning practitioners to better contextualize the work they do --- a vital aspect of any future responsible synthetic data work.

Further responsible dataset development frameworks that explicitly attend to the particulars of synthetic data, as well as tooling and practice that examines and makes transparent the provenance of synthetic data, are needed. As examples of this, we propose 'how could less risky synthetic data be produced?', as well as 'how could governance approach consent issues with synthetic data?' as important future research questions. In future work, we hope to follow the call of Denton et al.~\cite{dentonGenealogyMachineLearning2021}, contesting machine learning datasets and focusing on the “contingent, historical, and value-laden work practices of actual machine learning researchers” to better understand how the practice of using synthetic data is motivated, the contingent conditions that have lead to its common usage, and the norms and routines that surround it. In so doing, there is the opportunity to survey and better understand the use of synthetic data and create better tools and frameworks for both mitigating its potential for harmful power consolidation, as well as to envision how it can be used as a tool for taking power back~\cite{whitneyHCITacticsPolitics2021}.

\section{Conclusion}
In this paper, we build on prior responsible dataset development work by focusing on the under-explored impacts of synthetic data on dataset development. Synthetic data will continue to play an ever-increasing role in the training of machine learning systems as real-world data becomes harder to capture, and we must attend to language that paints it as a panacea. We show two examples of the risks of synthetic data, diversity-washing and consent circumvention, and discuss how it is a complicated tool, gravitationally prone to consolidation of power, but with potential for being used to enable participatory governance instead of squashing it.

\section{Researcher Ethics and Social Impact}
\subsection{Researcher Positionality Statement}
The first author is a white Latino AI researcher significantly influenced by their research, which has examined how policy and technical practice interplay and talk past each other, and how this dynamic affects those most likely to be harmed by AI systems. They worked in ML before moving into academia. The second author is a Black AI researcher with a variety of experiences in government, industry and now academia. They have access to the resources necessary to conduct their research, and recognize that that they have access to resources that many others do not. They strive to be conscious of their biases and to mitigate their impact on their work as much as possible. Both authors were motivated to write this paper by the realization that the risks of a commonplace technical practice were under-explored when discussing a real world example (detailed in this paper), and hoped to provide a starting point for understanding how using synthetic data could go wrong. Both researchers are based in the U.S., and that heavily influences both the examples they draw upon to show risks, the harms that they find salient, and the Overton window through which they view the world.

\subsection{Ethical Considerations Statement}
This work focuses on illustrating risks through the analysis and description of public-facing information and prior work through a new lens. As this is an example of RAI work that is focused on human impact but that does not involve study participants or create or deploy new technology, the main ethical consideration is in how this prior work is presented, where we actively attempted to avoid falling into some of the same traps we discuss --- we do not wish to make the technology seem inevitable or help to legitimize it, while we also do not want to forestall the opportunity for the risks we present to be mitigated and it to be used in participatory and ethical manners. We believe that the FRT evaluation example provided in this paper, created to assist in preventing unfounded FRT claims being used in the criminal justice setting, necessitated the creation of these datasets and the usage of synthetic data, but that is far from an ever-present conclusion.

\subsection{Adverse Impact Statement}
The largest adverse impact we are wary of is that these risks could be taken as playbooks --- we hope that nobody comes away thinking that there is opportunity to take advantage of them. We think that by making them public we are doing more of a good, as these examples demonstrate that the potential for these things already exists, and active exploration and research focused on mitigating and re-directing potential is the best way forward. We also see that this work could potentially draw scrutiny to legitimate uses of synthetic data, but hope that any added friction there is worth preventing potential malpractice.

\begin{acks}
This research was supported by UC Berkeley's Center for Long Term Cybersecurity AI Policy Hub, where the first author conducted research as a Fellow. This research was also supported by the National Science Foundation Graduate Research Fellowship under Grant No. DGE 1752814. Any opinion, findings, and conclusions or recommendations expressed in this material are those of the authors(s) and do not necessarily reflect the views of the National Science Foundation or of the CLTC. We thank those who helped us workshop and develop this material, specifically including Professor Deirdre Mulligan, Professor Jenna Burrell, Professor Hany Farid, Professor Brandie Nonnecke, Jessica Newman, Lauren Chambers and Seyi Olojo.
\end{acks}

\bibliographystyle{ACM-Reference-Format}
\bibliography{facct24}

\end{document}